\documentclass[aps,prl,twocolumn,showpacs,groupedaddress,nofootinbib]{revtex4}  
\usepackage{graphicx}  
\usepackage{dcolumn}   
\usepackage{bm}        
\usepackage{amssymb}   
\usepackage{slashed}
\usepackage{color}
\usepackage{lipsum}
\usepackage{textcomp}
\usepackage{appendix}
\begin{document}	
 \title{Inclusive weak-annihilation decays and lifetimes of beauty-charmed baryons}	
\author{Guo-He Yang$^{1}$\footnote{yghhust@hust.edu.cn}, En-Pei Liang$^{1}$\footnote{Enpei.Liang@outlook.com, Corresponding author}, Qin Qin$^{1}$\footnote{ qqin@hust.edu.cn, Corresponding author }, Kang-Kang Shao$^{2}$\footnote{shaokk18@lzu.edu.cn} }

\address{
$^1$School of Physics, Huazhong University of Science and Technology, Wuhan 430074, China \\$^2$School of Nuclear Science and Technology,  Lanzhou University, Lanzhou 730000,  China}
\date{\today}
\providecommand{\keywords}[1]{\textbf{\textit{keywords---}} #1}	
\begin{abstract} 
Imbalanced beauty-charmed baryons $ \Xi_{bc}^{+,0} $ are of great significance to the development of heavy flavor physics. In this work, we study the inclusive weak-annihilation decays of $\Xi_{bc}^{+,0}$ and their contributions to the $\Xi_{bc}^{+,0}$ lifetimes. For the calculation of the inclusive $\Xi_{bc}^{+,0}\to X_{cs}$ decay width where $X_{cs}$ stands for the sum of the final states with charm number +1 and strange number -1, we work in the heavy diquark effective theory which provides us with a convenient technical tool to construct the operator product expansion. The $\Xi_{bc}^{+,0}$ is considered to be a superposition of two states with one containing a spin-0 $bc$ diquark and the other one containing a spin-1 $bc$ diquark. It is found that both the $\Xi_{bc}^{+,0}$ lifetimes and the $\Xi_{bc}^{+,0}\to X_{cs}$ branching ratios are very sensitive to the $bc$ spin in $\Xi_{bc}^{+,0}$. The $\Xi_{bc}^{+,0}\to X_{cd}$ results are also presented. As $\Xi_{bc}^+$ has a longer lifetime than $\Xi_{bc}^{0}$ and bigger branching ratios of similar decay channels, the exclusive decays $\Xi_{bc}^+\to D^{(*)+}\Lambda$, $\Xi_{bc}^+\to \Lambda_c^+\bar{K}^{(*)0}$,  $\Xi_{bc}^+\to D^{(*)+}K^-p$, and $\Xi_{bc}^+\to D^{0}\bar{K}^{(*)0}p$ are more promising for experimental searches of $\Xi_{bc}^{+,0}$ at the LHC comparing with exclusive decay $\Xi_{b c} ^{0} \to D^0pK^-$.
\end{abstract}
\maketitle

\section{Introduction}

Doubly heavy hadrons provide a new platform to decipher the strong interaction. The discovery of the first doubly charmed baryon $\Xi_{c c} ^{++}$~\cite{LHCb:2017iph} has motivated many further studies. The LHCb collaboration has precisely measured its lifetime~\cite{LHCb:2018zpl}, mass~\cite{LHCb:2019epo} and production~\cite{LHCb:2019qed}. Besides the discovery channel proposed by~\cite{Yu:2017zst}, $\Xi_{cc}^{++}$ has also been searched via other channels~\cite{LHCb:2018pcs,LHCb:2019ybf,LHCb:2022rpd}. Theoretically, the $\Xi_{cc} ^{++}$ decay properties have been studied extensively by {\it e.g.}~\cite{Wang:2017mqp,Hu:2017dzi,Zhao:2018mrg,Xing:2018lre,Shi:2019hbf,Shi:2019fph,Xing:2021enr}. Experimental efforts have also been put into searching for the flavor SU(3) partners of $\Xi_{cc} ^{++}$~\cite{LHCb:2021rkb,LHCb:2021eaf}, but none of them has been discovered yet. On the other hand, the experimental discovery potential of doubly charmed tetraquarks has been theoretically studied in~\cite{Qin:2020zlg}, and thereafter the first of them, $T_{cc}^+$, was discovered at the LHCb~\cite{LHCb:2021vvq, LHCb:2021auc}. 

Unlike doubly charmed hadrons, the beauty-charmed baryons $\Xi_{bc}^{+,0}$ contains an imbalanced heavy quark pair, resulting in diverse features and deserving special attention. Experimental searches for the beauty-charmed baryons have been performed via the exclusive channels $ \Xi_{b c} ^{0} \to D^0pK^-$~\cite{LHCb:2020iko} and $ \Xi_{b c} ^{0} \to \Xi_{c}^+\pi^-$~\cite{LHCb:2021xba} however no significant signal was found. Very recently, the LHCb observed two peaking structures using $\Xi_{b c}^{+}\to J / \psi \Xi_{c}^{+}$~\cite{LHCb:2022fbu} with a local (global) significance of 4.3 (2.8) and 4.1 (2.4) standard deviations at masses of 6571 MeV and 6694 MeV, respectively, which might be very close to the discovery. Besides, it was proposed by~\cite{Qin:2021wyh} that the inclusive decay $\Xi_{bc} \to \Xi_{cc}^{++}+X$ may serve as the discovery channel of $\Xi_{bc}^{+,0}$ by making use of the displacement information of $\Xi_{cc}^{++}$, since plenty of the $\Xi_{bc}$ baryons will be produced at the LHC~\cite{Ali:2018ifm,Ali:2018xfq}.

We study in this work the inclusive weak-annihilation decays of $\Xi_{bc}^{+,0}$, especially $\Xi_{bc}^{+,0}\to X_{cs}$, where $X_{cs}$ stands for the sum of the final states with charm number +1 and strange number -1. One example exclusive channel contributing to this inclusive decay is $\Xi_{bc}^{0}\to D^0 p K^-$, which has been used for $\Xi_{bc}^{0}$ search~\cite{LHCb:2020iko}. The inclusive decay rate can help evaluate whether the corresponding exclusive channels such as $\Xi_{bc} ^{0} \to D^0pK^-$ and $\Xi_{bc}^+\to D^{(*)+}\Lambda$, $\Xi_{bc}^+\to \Lambda_c^+\bar{K}^{(*)0}$,  $\Xi_{bc}^+\to D^{(*)+}K^-p$, and $\Xi_{bc}^+\to D^{0}\bar{K}^{(*)0}p$ are potential discovery channels of the beauty-charmed baryons. From the theoretical point of view, this process has a simple structure that the heavy diquark part can be factorized out at the lowest order of the strong coupling $\alpha_s$. Therefore, the calculation with lower model dependencies can make the inclusive  decay $\Xi_{bc}^{+,0}\to X_{cs}$ as a test of the heavy diquark effective theory (HDET)~\cite{Shi:2020qde, Qin:2021wyh}. 

Previously, the inclusive $\Xi_{bc}^{+,0}\to X_{cs}$ decay has been studied in~\cite{Kiselev:1999kh,Kiselev:2001fw} where the lifetimes of $\Xi_{bc}^{+,0}$ were investigated under the assumption that the beauty and charm quarks form a scalar $bc$ diquark inside $\Xi_{bc}^{+,0}$. On the other hand, the $bc$ diquark was treated as an axial-vector state in~\cite{Cheng:2019sxr} and the results for both the $\Xi_{bc}^{+,0}\to X_{cs}$ decay widths and the $\Xi_{bc}^{+,0}$ lifetimes are very different. We take into account the possibility that the $bc$ diquark in $\Xi_{bc}^{+,0}$ is a superposition of the scalar and axial-vector states, $\mathcal{S}_{b c}$ and $\mathcal{X}_{b c}$~\cite{Roberts:2008wq}. It confirms that the scalar and axial-vector contributions are extremely different, so the $\Xi_{bc}^{+,0}$ lifetimes and the $\Xi_{bc}^{+,0}\to X_{cs}$ decay rates can be used to determine the diquark constituent. This study also improves the calculation in the following aspects: (i) We revisit the inclusive channel more systematically in the heavy diquark effective theory~\cite{Shi:2020qde, Qin:2021wyh}, including power corrections of $1/m_{bc}^n$, where $m_{bc}$ is the $bc$ diquark mass. (ii) We consider the contribution from the QCD penguin operators. The inclusive $\Xi_{bc}^{+,0}\to X_{cd}$ results are obtained by the corresponding Cabbibo-Kobayashi-Maskawa (CKM) matrix element replacement.

The rest of the paper is organized as follows. In section 2, we introduce the HDET and perform the matching for involved operators. In section 3, the inclusive decay rate is formulated in the framework of operator product expansion, and it turns out that the result is restricted by the reparametrization invariance. The numerical results and relevant phenomenological discussions are given in section 4. We conclude with section 5.

\section{Heavy Diquark Effective Theory}

The $\Xi_{bc}$, as a doubly heavy baryon, has a heavy diquark - light quark structure, like a “double-star” core surrounded by a light “planet”. The effective distance between the two heavy quarks is much smaller than that between any one of them and the light quark, so the two heavy quarks can be regarded as a point-like heavy diquark if the physics above the QCD scale $\Lambda_{\rm{QCD}}$ is integrated out. (See \cite{Yin:2019bxe,Yin:2021uom} for a different description.) In the heavy diquark limit as its mass $ m_{QQ'} \to \infty $, the heavy diquark system can be treated as a static $\bar{3}$ color source, playing the same role as a heavy anti-quark in a heavy meson~\cite{Georgi:1990ak}. 

The spin of the heavy diquark $ J_{bc} $ is a good quantum number in the heavy diquark limit. Therefore, we can use $J_{bc}$ to label different diquark states, spin-0 scalar $\mathcal{S}_{bc}$ and spin-1 axial-vector $\mathcal{X}_{b c}$. The ground state $\Xi_{b c}$ is spin-1/2, so it can be composed of either $\mathcal{S}_{bc}$ or $\mathcal{X}_{b c}$ with light degrees of freedom $q$ with $J_{q}= 1/2$. In principle, $\Xi_{b c}$ is a superposition of the two states $\Xi_{\mathcal{S}}(\mathcal{S}_{bc}q)$ and $\Xi_{\mathcal{X}}(\mathcal{X}_{b c}q)$, formulated as 
\begin{eqnarray}
	\left|\Xi_{b c}^{+,0}\right\rangle= \cos\theta\left|\Xi_{\mathcal{X}} \right\rangle +\sin\theta e^{i\phi} \left|\Xi_{\mathcal{S}}\right\rangle, 
\end{eqnarray}
where $\theta$ is the mixing angle and $\phi$ is the relative phase.

Both the scalar $\mathcal{S}_{bc}$ and axial-vector $\mathcal{X}_{bc}$ can be described systematically in the HDET, with the Lagrangian given by~\cite{Shi:2020qde}
\begin{eqnarray}\label{eq:hdet}
\mathcal{L}_{\mathcal{S}}&= &i  \mathcal{S}_{v}^{\dagger} v \cdot D \mathcal{S}_{v}-\frac{1}{2m_\mathcal{S}} \mathcal{S}_{v}^{\dagger} D^{2} \mathcal{S}_{v} + \mathcal{O}(\frac{1}{m_\mathcal{S}^2}) \;, \nonumber \\
\mathcal{L}_{\mathcal{X}}&=&-i \mathcal{X}_{v \mu}^{\dagger} v \cdot D \mathcal{X}_{v}^{\mu}+ \frac{1}{2m_\mathcal{X}} \mathcal{X}_{v \mu}^{\dagger} D^{2} \mathcal{X}_{v}^{\mu} \nonumber\\
&~&  +\frac{ig}{2m_\mathcal{X}}\mathcal{X}_{v \mu}^{\dagger}\bar{G}^{\mu \nu}\mathcal{X}_{v \nu} +\mathcal{O}(\frac{1}{m_\mathcal{X}^2}) \; ,
\end{eqnarray}
where $D_{\mu}=\partial_{\mu}-i g A_{\mu}^{a} \overline{t^{a}}$, $g$ is the effective coupling constant between the diquark and the gluon, 
$ \overline{t^{a}} $ is the $\bar{3}$ generator of the color $ SU(3)$ group, and the gluon tensor is defined by $\bar{G}_{\mu \nu}= \frac{-i}{g}[D_{\mu}, D_{\nu}]=G_{\mu \nu}^a \overline{t^{a}}$. The $v$-subscripted scalar and axial-vector fields are related to the original fields by the definition 
\begin{eqnarray}  
\mathcal{S}(x)&=&\exp [-i m_\mathcal{S}v \cdot x] \mathcal{S}_{v}(x)/\sqrt{2m_\mathcal{S}} \; ,\nonumber  \\
\mathcal{X}^{\mu}(x)&=& \exp [-i m_\mathcal{X}v \cdot x] (\mathcal{X}_{v}^{\mu}(x)+Y_{v}^{\mu}(x))/\sqrt{2m_\mathcal{X}} \; ,\ \ \ 
\end{eqnarray}
with the large momentum $mv$ subtracted, where the reference velocity $v$ is often chosen to be the baryon velocity. The static part $ \mathcal{X}_{v}^{\mu} $ satisfies $ v\cdot  \mathcal{X}_v = 0$ and the residual part $ Y_{v}^{\mu} $ is suppressed  by powers of $\Lambda_{\rm{QCD}}/m_\mathcal{X}$. The leading terms in (\ref{eq:hdet}) have the heavy diquark spin-flavor symmetry. Compared to~\cite{Shi:2020qde}, the fields are different by a factor of mass square root to make the heavy diquark flavor symmetry manifest. 

The effective Hamiltonian involved in the $bc\to cs$ processes is~\cite{Buchalla:1995vs}
\begin{eqnarray}\label{eq:effH}
	&\mathcal{H}_{\text{eff}}=\frac{4 G_{F}}{\sqrt{2}} V_{c s}^{*} V_{c b}\sum_{i=1}^{6}C_{i}O_{i} \\
&O_{1}=\bar{c}_{\lambda} \gamma^{\mu} P_{L} b_{\rho}\bar{s}_{\rho} \gamma_{\mu} P_{L} c_{\lambda}\; , \ O_{2}=\bar{c} \gamma^{\mu} P_{L} b\; \bar{s}\gamma_{\mu} P_{L} c \; , \nonumber\\
	&O_{3}=\bar{s} \gamma^{\mu} P_{L} b \; \bar{c} \gamma_{\mu} P_{L} c\; , \ O_{4}=\bar{s}_{\lambda} \gamma^{\mu} P_{L} b_{\rho}\bar{c}_{\rho} \gamma_{\mu} P_{L} c_{\lambda} \; , \nonumber\\
	&O_{5}=\bar{s} \gamma^{\mu} P_{L} b\; \bar{c} \gamma_{\mu} P_{R} c \; , \ O_{6}=\bar{s}_{\lambda} \gamma^{\mu} P_{L} b_{\rho}\bar{c}_{\rho} \gamma_{\mu} P_{R} c_{\lambda} \; , \nonumber 
\end{eqnarray}
where $P_{L,R}$ are the left- and right-handed projectors, respectively. To match these operators to the diquark-quark operators, we write the scalar and axial-vector diquark states as~\cite{Shi:2020qde}
\begin{eqnarray}
	\left|\mathcal{S}_{bc}^{i}(v)\right\rangle=&\frac{\sqrt{ E_{\mathcal{S}}}}{4\sqrt{2m_{b}m_{c}}} \int \frac{d^{3} \mathbf{k}}{(2 \pi)^{3}} \phi^{*}(\mathbf{k}) \epsilon_{i j k}  \left[C \gamma_{5}(1+\slashed{v} )\right]_{\beta \gamma}^{\dagger} \nonumber \\
	& Q_{j \beta}^{\dagger}(v, \mathbf{k}) Q_{k \gamma}^{\prime \dagger}(v,-\mathbf{k})|0\rangle \;, \\	
	\left|\mathcal{X}_{bc}^{i}(v, \epsilon)\right\rangle=&\frac{-\sqrt{ E_{\mathcal{X}}}}{4\sqrt{2m_{b}m_{c}}} \int \frac{d^{3} \mathbf{k}}{(2 \pi)^{3}} \phi^{*}(\mathbf{k}) \epsilon_{i j k}  [C \slashed{\epsilon}(1+\slashed{v})]_{\beta \gamma}^{\dagger} \nonumber \\
	&Q_{j \beta}^{\dagger}(v, \mathbf{k}) Q_{k \gamma}^{\prime \dagger}(v,-\mathbf{k})|0\rangle \; ,
\end{eqnarray}
where the heavy quark operators are defined by 
\begin{eqnarray}
	Q(v, \mathbf{k})=\sum_{s} u^{s}(p) a_{\mathbf{p}}^{s}, \text { with } \mathbf{p} \equiv m_{Q} \mathbf{v}+\mathbf{k} \; . 
\end{eqnarray}	
The function $\phi(\mathbf{k})$ is the distribution of the relative momentum $\mathbf{k}$ of the quarks in the heavy diquark bound states, and $\mathbf{k}$ is typically small compared to the heavy quark masses. In the heavy quark limit, the distribution $\phi(\mathbf{k}) $ is concentrated at $\mathbf{k} = 0$.  The wave functions in the spinor space $\left[C \gamma_{5}(1+\slashed{v} )\right]_{\beta \gamma} $ and $\left[C \slashed{\epsilon}(1+\slashed{v} )\right]_{\beta \gamma}  $ are anti-symmetric and symmetric, respectively, with $\epsilon$ the polarization vector of axial-vector diquark field and $C\equiv i\gamma^0\gamma^2$. The antisymmetric tensor $\epsilon_{ijk}$ is the wave function in the color spaces with $i,j,k$ color indices. The projector $ (1+\slashed{ v}) $ approximately selects the particle component and eliminates the antiparticle component. The expressions are not well defined from the first principle of QCD, but they can give the correct matching results in the leading power of inverse heavy quark mass. 


After performing the match by calculating the $\mathcal{S}_{bc}\to cs$ and $\mathcal{X}_{bc}\to cs$ transition amplitudes at the lowest order of $\alpha_s$ and $\mathbf{k}$, we obtain the scalar diquark-quark interaction terms in the Hamiltonian. 
\begin{eqnarray}
\mathcal{H}_\mathcal{S}&\ni &A_{12}\epsilon_{ijk}\; \mathcal{S}^{i}\; \bar{c}_j P_{R}C\bar{s}^{T}_k
+ A_{34} \epsilon_{ijk }\; \mathcal{S}^{i}\; \bar{c}_j P_{R}C\bar{s}^{T}_k \nonumber \\
&~&+A_{56}/m_{bc}\; \epsilon_{ijk}\; (iD_\mu\mathcal{S}^{i}) \; \bar{c}_j \gamma^\mu P_{R}C\bar{s}^{T}_k \; , \\
	A_{12}&=&4\sqrt{m_{bc}}  G_{F}V_{c s}^{*} V_{c b}(C_{2}-C_{1}) \psi_{b c}(0) \; , \nonumber\\ 
	A_{34}&=&4\sqrt{m_{bc}} G_{F}V_{c s}^{*} V_{c b}(C_{4}-C_{3}) \psi_{b c}(0) \; , \nonumber\\ 
	A_{56}&=&2\sqrt{m_{bc}}  G_{F}V_{c s}^{*} V_{c b}(C_{5}-C_{6}) \psi_{b c}(0)   \; ,\nonumber 
\end{eqnarray}
and the axial-vector diquark-quark interaction term 
\begin{eqnarray}\label{eq:axial}
\mathcal{H}_\mathcal{X}&\ni &B_{5,6}\epsilon_{i jk }\bar{c}_{j}\gamma^{\mu}P_{R}C\bar{s}^{T}_k\mathcal{X}_{v,\mu}^{i} \; , \\
	B_{56}&=&-A_{56}  \; .\nonumber 
\end{eqnarray}
The $\psi_{b c}(0)$ is the Fourier transformed of $\phi(\mathbf{k})$ at the zero point. 


\section{Operator product expansion}

The decay rate of the inclusive $\Xi_{bc}^{+,0}\to X_{cs}$ channel can be expressed as 
\begin{eqnarray}
	\Gamma(\Xi_{bc}^{+,0} \to X_{cs}) &=& \sum_{X_{cs}} (2\pi)^4 \delta^{(4)}(p_\Xi-p_X) {|\langle X_{cs}| \mathcal{H}_I|\Xi_{bc}^{+,0}\rangle  |^2 \over 2m_{\Xi}} \nonumber \\
	&=& {\text{Im} \langle \Xi_{bc}^{+,0}| \mathcal{T} | \Xi_{bc}^{+,0} \rangle\over m_{\Xi} } \; ,
\end{eqnarray}
where the operator product $\mathcal{T}$ is given by 
\begin{eqnarray}
	\mathcal{T} = i\int d^4x\; T\{ \mathcal{H}_I^\dagger(x) \mathcal{H}_I(0) \} \; , \mathcal{H}_I =  \mathcal{H}_\mathcal{S} +\mathcal{H}_\mathcal{X} \; , 
\end{eqnarray}
by making use of the optical theorem. The nonlocal operator product $\mathcal{T}$ is further expanded as 
\begin{eqnarray}\label{eq:ope}
	\mathcal{T} =  \sum_i F_i Q^{(n)}_i \; ,
\end{eqnarray}
with the technique called local operator product expansion (OPE), where the $Q^{(n)}_i$ are local operators with dimension $n$ and $F_i$ are the corresponding coefficients. Contributions from higher dimensional operators are suppressed, in this case, by powers of $\Lambda_{\text{QCD}}/m_{bc}$.

\begin{figure}[!ht]
	\centering
	\includegraphics[width=0.6\linewidth]{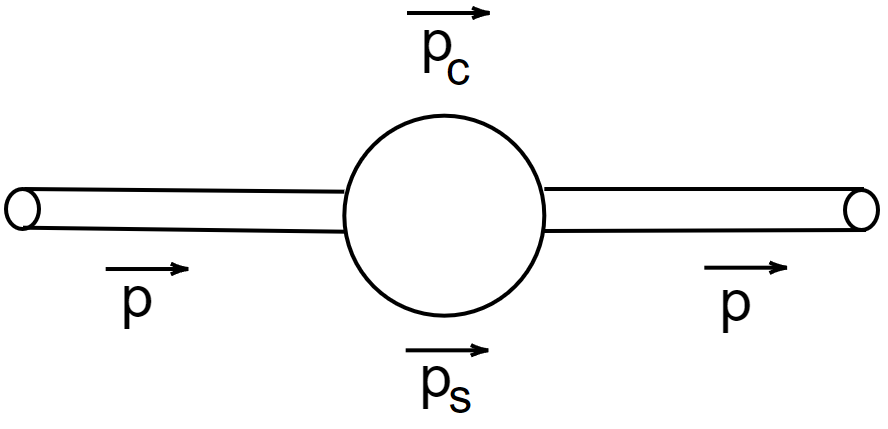}
	\caption{The leading order diagram for the diquark-to-diquark amplitudes.}
	\label{fig:1}
\end{figure}

The OPE is performed by requiring that the quark-diquark matrix elements of the left-handed side and the right-handed side of (\ref{eq:ope}) match each other. Starting with the scalar diquark, the matching is performed by calculating the $ \mathcal{S}_{bc} \to cs \to \mathcal{S}_{bc}$ amplitude at the least order of $\alpha_s$, as displayed in FIG.~\ref{fig:1}. The imaginary part of the amplitude is calculated as 
\begin{eqnarray}  \label{eq:SS}
&&{\rm{Im}}\;\mathcal{M}(S_{bc }\to S_{bc }) =A\frac{(p^2-m_c^2)^2 }{8 \pi p^2}\;,\\
&& A=|A_{12}+A_{34}+{m_{c}^2\over m_{bc}^2}A_{56}|^2\;,			\nonumber 
\end{eqnarray}
where the diquark has momentum $ p=m_{bc}v+k $, with the residual momentum $k\sim \Lambda_{\text{QCD}}$, and the expansion is performed by powers of $k$. 
The $k^0$ term in the expansion of (\ref{eq:SS}) is reproduced by the dimension-3 operator with the coefficient 
\begin{eqnarray}  
	Q^{s(3)}_1 = S^{\dagger}_{v}S_{v}\; , \;  F^s_1=\frac{A(m_{bc}^2-m_c^2)^2 }{16 \pi m_{bc}^3} \; . 
\end{eqnarray}
The $k^1$ term and the $k^2$ term correspond to the dimension-4 operator $S^{\dagger}_{v} (i v\cdot D)S_{v}$ and the dimension-5 operator $S_{v}^{\dagger}(iD)^2S_{v}$, respectively. These two operators are related by the equation of motion $[i v \cdot D+{(iD)^2}/(2m_{bc})]S_v=0$. 
Taking their coefficients ${A(m_{bc}^4-m_c^4)  }/({8 \pi m_{bc}^4})$ and ${A(m_{bc}^4-m_c^4)  }/({16 \pi m_{bc}^5})$, it turns out that their contribution exactly cancel each other. This is actually protected by the reparametrization invariance proved by~\cite{Mannel:2018mqv} \footnote{The decomposition of $p$ into $v$ and $k$ is not unique. A small change in the four-velocity of the order of $\Lambda_{\mathrm{QCD}} / m_{bc}$ can be compensated by a change in the residual momentum: $v  \rightarrow v+\delta v, k  \rightarrow k-m \delta v$.}. Therefore, up to $\mathcal{O}(\Lambda_{\text{QCD}}^2/m_{bc}^2)$, only the matrix element of $Q^{(3)}_1$ needs to be evaluated. (The chromomagnetic operator does not exist in the scalar case.) It can be parameterized as~\cite{Mannel:2018mqv}
\begin{eqnarray}  
	&&\left \langle \Xi_S|S^{\dagger}_{v}S_{v}|\Xi_S\right\rangle \equiv  2m_{\Xi} \mu_3 =  2m_{\Xi} (1- {\mu_\pi^2\over 2 m_{bc}^2}) \; ,	\\
	&&\text{with}\ \left \langle \Xi_S|S^{\dagger}_{v} (iD)^2S_{v}|\Xi_S\right\rangle = -2m_{\Xi} \mu_\pi^2 \; .
\end{eqnarray}
The nonperturbative parameter $\mu_\pi^2$ takes the value $ 0.43\pm0.24 $~\cite{Bernlochner:2022ucr} in the numerical analysis, which is extracted from the inclusive $B$ meson decay. We expect that the $\mu_\pi^2$ values for $\Xi_{bc}^{+,0}$ and $B$ mesons are the same owing to the heavy quark-diquark symmetry.

The imaginary part of the $\mathcal{X}_{bc} \to cs \to \mathcal{X}_{bc}$ amplitude corresponding to FIG.~\ref{fig:1} is
\begin{eqnarray}  
	&&{\rm Im}\;\mathcal{M}_{5,6}(\mathcal{X}_{bc }\to \mathcal{X}_{bc }) =B\frac{(p^2-m_c^2)^2 }{8 \pi p^2}  \; ,\\
&&	B=|B_{56}|^2 (2+{m_c^2\over m_{bc}^2}) \; . \nonumber
\end{eqnarray}
Because the whole amplitude is suppressed by the penguin Wilson coefficients $(C_5-C_6)^2$, we only perform the OPE to the leading power, 
\begin{eqnarray}  
&&\mathcal{T} \ni  F_1^a Q_1^{a(3)} \; ,  \\
&&Q_1^{a(3)} = \mathcal{X}_{v}^{\mu\dagger}\mathcal{X}_{v\mu}\; , \;  F^a_1=\frac{B(m_{bc}^2-m_c^2)^2 }{16 \pi m_{bc}^3}  \; .\nonumber
\end{eqnarray}
Analogous to the scalar case, the hadronic matrix element of $Q_1^{a(3)}$ is 
\begin{eqnarray}  
\left \langle \Xi_{\mathcal{X} }| \mathcal{X}_{v}^{\mu\dagger}\mathcal{X}_{v\mu} |\Xi_{\mathcal{X} }\right\rangle \equiv  2m_{\Xi} \mu_3 \; .
\end{eqnarray}
The $S_{bc }\to cs\to X_{bc}$ amplitude at the leading order vanishes because of angular momentum conservation, so there would be no interference between the scalar and axial-vector constituents in the $\Xi_{bc}^{+,0}$ baryons.

\section{Numerical analysis}

If the $bc$ diquark in $\Xi_{bc}$ is purely scalar, the leading-order inclusive decay width is given by 
\begin{eqnarray}\label{eq:gammas}
	&&\Gamma (\Xi_{S} \to  X_{cs})\nonumber\\
	&=&\frac{2G_{F}^2}{ \pi m_{bc}^2} [C_{2}-C_{1}+C_{4}-C_{3}+\frac{m_c}{2m_{bc}}(C_{5}-C_{6})]^2 \nonumber\\
	&&|V_{c s}^{*} V_{c b}|^2|\psi_{b c}(0) |^2	(m^2_{bc}-m^2_c)^2 (1- {\mu_\pi^2\over 2 m_{bc}^2})  \nonumber \\
&	\simeq& (5.7\pm2.1)\times 10^{-13} \; \text{GeV}\; .
\end{eqnarray}
where the Fermi constant, quark masses, and the CKM matrix elements take values from~\cite{ParticleDataGroup:2020ssz} and the Wilson coefficients take values from~\cite{Lu:2000em}. The uncertainty arise from the variation of the quark masses, the Wilson coefficients, the CKM matrix elements, the diquark wave function at the origin $\psi_{b c}(0)=\frac{0.87 \pm0.09}{\sqrt{4\pi}} \mathrm{GeV}^{3 / 2}$~\cite{Qin:2021wyh}, the nonperturbative parameter $\mu_\pi^2=0.43\pm0.24$~\cite{Bernlochner:2022ucr} and we have include additional $\sim30\%$ uncertainty from possible power corrections~\cite{Zhu:2017lqu}. It can be checked that the leading power result recovers the free diquark decay rate. The $\Xi_{S} \to X_{cd}$ result can be obtained if we replace $V_{cs}$ with $V_{cd}$, and it reads $\Gamma(\Xi_{S} \to  X_{cd})= (2.7\pm1.0)\times 10^{-14}$ GeV. This procedure also applies to the rest results. 

\begin{figure}[!ht]
	\centering
	\includegraphics[width=0.69\linewidth]{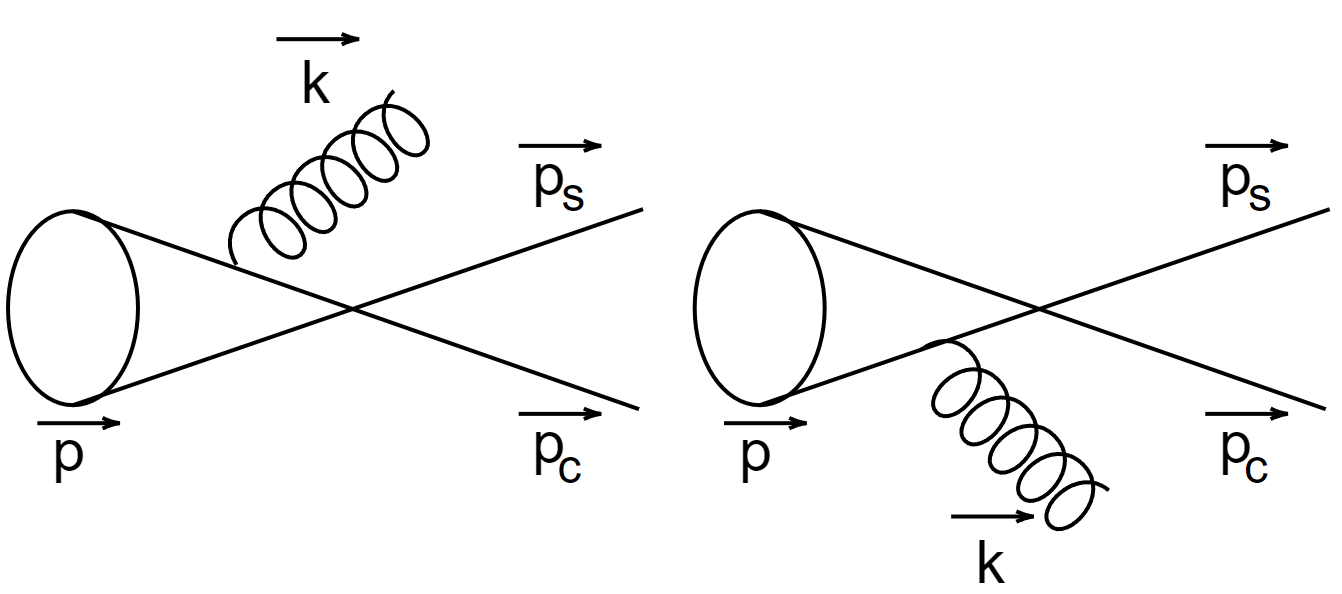}
	\caption{The next-to-leading order Feynman diagrams contributing to $\Xi_{\mathcal{X}} \to  X_{cs}$ with a real gluon emitted from the diquark.}
	\label{fig:2}
\end{figure}

If the diquark in $\Xi_{bc}$ is purely axial-vector, the leading order inclusive decay width is given by 
\begin{eqnarray}\label{eq:gammaa}
&	&\Gamma_1(\Xi_{\mathcal{X}} \to  X_{cs})\nonumber\\
&	=&\frac{G_{F}^2}{6 \pi m^2_{bc}} |V_{c s}^{*} V_{c b}|^2(C_{5}-C_{6})^2 |\psi_{b c}(0) |^2
	\nonumber\\
&	&	(m^2_{bc}-m^2_c)^2 (2+\frac{m_c^2}{m_{bc}^2}) (1-{\mu_\pi^2\over 2 m_{bc}^2})\nonumber\\
&\simeq& (9.0\pm 3.4)\times 10^{-17}\; \text{GeV} \; ,
\end{eqnarray}
where the uncertainties are from the same sources as (\ref{eq:gammas}). The leading power result again recovers the free diquark decay rate.
Practically, the leading-order result for the $\Xi_{\mathcal{X}}$ decay rate is highly suppressed by the Wilson coefficients, such that the next-to-leading-order (NLO) contributions might be larger. We consider the NLO correction with a real gluon emitted from the diquark, as shown in FIG.2. Taking into account the tree operators in (\ref{eq:effH}), we obtain their NLO contribution 
\begin{eqnarray}
	&&\Gamma_2(\Xi_{\mathcal{X}} \to  X_{cs})\nonumber	\\
	&=&\frac{\alpha_{s}}{36\pi^{2}m_{bc}^4}[(C_{1}^2+C_{2}^2)(x^2-xy+y^2)+C_{1}C_{2}(x^2-\nonumber \\
	&&4xy+y^2)]G_{F}^{2}\left|V_{c s}^{*} V_{c b}\right|^{2}|\psi_{b c}(0) |^2\nonumber \\
	&&\int^{m_{b}^2}_{0} d s_{12} \bigg[
	\frac{1}{2} \left(-m_c^2-s_{12}+m_{bc}^2\right) \nonumber \\
	&&\sqrt{-2 m_c^2 \left(s_{12}+m_{bc}^2\right)+\left(m_{bc}^2-s_{12}\right)^2+m_c^4} \bigg]	 \nonumber\\
	&\simeq&(4.7\pm1.8)\times10^{-16} \; \text{GeV} \; ,
\end{eqnarray}
where $x=m_{bc}/m_b,y=m_{bc}/m_c$. The strong interaction constant $\alpha_s$ takes values from~\cite{ParticleDataGroup:2020ssz}. The uncertainties are from the same sources as (\ref{eq:gammas}). Indeed, it is larger than the leading-order contribution (\ref{eq:gammaa}) owing to enhanced Wilson coefficients. The virtual correction requires systematical renormalization of the HDET, which is beyond the scope of this study and is left for future works. To estimate the size of the virtual correction, we refer to~\cite{Cox:1985dj}, in which the $B\to J/\psi + X$ decay is studied with the same weak interaction vertex as our process. Their results show that the vertex correction is several times smaller than the hard bremsstrahlung correction, and approximately 10\% of the leading-order contribution. Therefore, we expect that the virtual correction might cause another uncertainty $\sim10\% $. As for multi-particle contributions such as a quark-anti-quark pair emission $\mathcal{X}_{bc}\to csq\bar{q}$, they are highly suppressed by $\alpha_s^2$ and the phase space~\cite{Huber:2018gii} and are thus neglected. 

Combining these two contributions, the inclusive $\Xi_{\mathcal{X}}$ decay rate is calculated to be 
\begin{eqnarray}
	&&\Gamma (\Xi_{\mathcal{X}}  \to  X_{cs})\simeq(5.6\pm2.2 )\times 10^{-16} \; \text{GeV} \; .
\end{eqnarray}
The corresponding $X_{cd}$ result is $\Gamma(\Xi_{\mathcal{X}}  \to  X_{cd}) =(2.7\pm1.0)\times 10^{-17}$  GeV.

In a general case, $\Xi_{b c}$ might be a superposition of $\Xi_{\mathcal{S}}$ and $\Xi_{\mathcal{X}}$~\cite{Roberts:2008wq,Weng:2018mmf,Roberts:2007ni}. Then, the inclusive $\Xi_{b c}^{+,0} \to X_{cs}$ decay width is expressed as
 \begin{eqnarray}  
&& 	\Gamma(\Xi_{b c}^{+,0} \to  X_{cs})\nonumber\\
 &=&\sin^2\theta\;\Gamma(\Xi_{S} \to  X_{cs})+\cos^2\theta\; \Gamma(\Xi_{\mathcal{X}} \to  X_{cs}) \; .
 \end{eqnarray}
Note that the interference between $\Xi_{\mathcal{S}}$ and $\Xi_{\mathcal{X}}$ vanishes because of angular momentum conservation. 

\begin{figure}[!ht]
	\centering
	\includegraphics[width=0.8\linewidth]{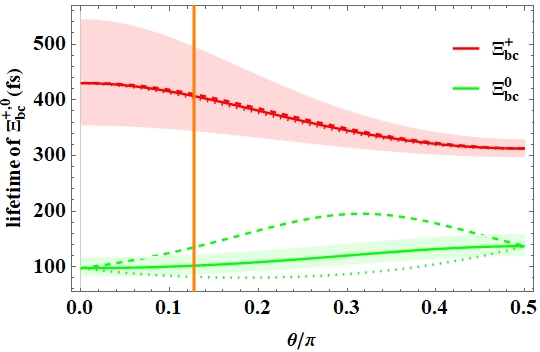}
	\caption{The lifetimes of $ \Xi_{b c}^{+,0} $ as functions of
		the mixing angle $ \theta $. The solid, dotted and dashed curves correspond to the choices of the phase angle $\cos\phi=0$, $\cos\phi=1$, and $\cos\phi=-1$, respectively. The vertical line corresponds to the reference value $\sin\theta$ = 0.39~\cite{Roberts:2008wq}.}
	\label{fig:lifetime}
\end{figure}

\begin{figure}[!ht]
	\centering
	\includegraphics[width=0.8\linewidth]{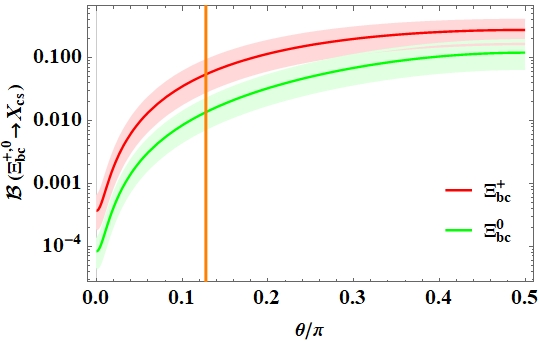}
	\caption{The branching ratios $\mathcal{B}(\Xi_{b c}^{+,0}\to  X_{cs})$ as functions of the mixing angle $\theta$ with $\cos\phi=0$, where the vertical line corresponds to the reference value $\sin\theta$ = 0.39~\cite{Roberts:2008wq}.} \label{fig:Br(bc+0)Log}
\end{figure}

Besides the $bc$ annihilation, the $bq$ and $cq$ annihilation contributions to the $\Xi_{bc}^{+,0}$ lifetimes also depend on the $bc$ diquark spin~\cite{Cheng:2019sxr,Kiselev:1999kh,Kiselev:2001fw}. The corresponding analytical results are given in Appendix. Accepting the hadronic inputs in~\cite{Cheng:2019sxr}, we numerically calculate these contributions with $\Xi_{bc}^{+,0}$ either being $\Xi_\mathcal{X}$ or $\Xi_\mathcal{S}$, as listed in TABLE~\ref{tb:gamma}, where the other contributions independent on the $bc$ spin are also listed. Then, the mixing angle-dependent results for the total decay widths of $\Xi_{bc}^{+,0}$ are given by 
\begin{eqnarray}  
& 	\Gamma(\Xi_{b c}^{+,0}) =&\sin^2\theta\; \Gamma_{\mathcal{S}}^{\text {an}}+\cos^2\theta\;\Gamma_{\mathcal{X}}^{\text {an}}\nonumber\\
&& + 2\sin\theta\cos\theta\cos\phi\Gamma_{\mathcal{SX}}^{\text {an}}+\Gamma^{\text {other}}\; , 
\end{eqnarray}
which can be translated to lifetimes shown in FIG.~\ref{fig:lifetime}. With the lifetimes, the branching ratios of the inclusive $\Xi_{b c}^{+,0} \to  X_{cs}$ decays are obtained, as shown in FIG.~\ref{fig:Br(bc+0)Log}. 

\begin{table} 
\begin{tabular}{lcccc}
	\hline \hline & $\Gamma_{\mathcal{S}}^{\text {an}}$  & $ \quad\Gamma_{\mathcal{X}}^{\text {an}}$&$ \Gamma_{\mathcal{SX}}^{\text {an}} $&$\Gamma^{\text {other}}$ \\
	\hline$\Xi_{b c}^{0}$ &$1.87\pm0.12$ &$3.81\pm0.30$&$2.20\pm0.17$ &$  1.50\pm 0.31 $ \\
	$\Xi_{b c}^{+}$ & $0.61\pm0.22$ & $0.035\pm0.008 $&$ 0.020\pm0.005 $  &$2.94\pm0.73$\\	
	\hline \hline
\end{tabular}
\caption{Various contributions to the total decay widths of $\Xi_{b c}^{+,0}$ in units of $10^{-12} \mathrm{GeV}$, including the \{$bc$,$bu$\} and \{$bc$,$cd$\} weak-annihilation contributions to $\Xi_{b c}^{+}$ and $\Xi_{b c}^{0}$, respectively. The subscript $\mathcal{S}$ and $\mathcal{X}$ represent the cases with the $bc$ diquark being scalar or axial-vector. The other contributions $\Gamma^{\text{other}}$ contain the spectator decay contribution $\Gamma_{}^{\text {dec}}$ and the Pauli interference $\Gamma^{\text {int}}$~\cite{Cheng:2019sxr}, which are independent on the diquark spin.}\label{tb:gamma}
\end{table} 

The mixing angle has been studied by many previous studies, {\it e.g.}~{\cite{Roberts:2008wq, Weng:2018mmf, Roberts:2007ni}}. As the predictions for the mixing angle $\theta$ are very diverse, we vary $\theta$ from 0 to $\pi/2$ and display the corresponding results in FIG.~\ref{fig:lifetime}. The lifetime of $ \Xi_{b c}^{0}) $ is heavily influenced by the phase angle, and the lifetime of $ \Xi_{b c}^{+}) $ has almost no effect. We also take $\cos\phi=0$, $\sin\theta \simeq0.39$ and $ \cos\theta\simeq0.92$~\cite{Roberts:2008wq} as a reference, and obtain the corresponding $\Xi_{b c}^{+,0} \to X_{cs}$ branching ratios,  
\begin{eqnarray}  
	\mathcal{B}(\Xi_{b c}^{0}\to  X_{cs}) \simeq (1.4^{+0.9}_{-0.6})\%  \; ,\\
	\mathcal{B}(\Xi_{b c}^{+ }\to  X_{cs}) \simeq (5.4^{+3.8}_{-2.6})\% \; .
\end{eqnarray}
The branching ratio of $ \Xi_{bc}^{+,0} \to  X_{cd} $ are $  (0.26^{+0.18}_{-0.10})\%$ and $  (0.07^{+0.04}_{-0.02})\%$, respectively.

It is observed that the lifetimes and the inclusive branching ratios are very sensitive to the mixing angle, so their measurements in future experiments can be used to determine whether the diquark in the ground state $\Xi_{bc}^{+,0}$ is scalar or axial-vector, or how they mix each other.

As for the $\Xi_{bc}^{+,0}$ search, it was estimated in~\cite{Kiselev:2001fw} that the $\mathcal{O}(10\%)$ inclusive $\Xi_{bc}^{+,0}\to X_{cs}$ decay branching ratio would indicate the typical exclusive channels $\Xi_{bc}\to D^{(*)}K^{(*)}p$ have branching ratios of several permil. With some estimated theoretical and experimental inputs, the result of~\cite{LHCb:2020iko} can be converted to the upper limit $\mathcal{B}(\Xi^0_{bc}\to D^{0}K^{-}p)\lesssim 0.3\%$ assuming that the $\Xi^0_{bc}$ lifetime is 100 fs. The experimental upper limit is very close to the theoretical estimation, so it is quite hopeful to discover $\Xi_{bc}$ by this channel with more data collected. In fact, considering the lifetime hierarchy between $\Xi_{bc}^0$ and $\Xi_{bc}^+$, searches via an analogous $\Xi_{bc}^+$ decay channel would be more promising. Taking the reference with $\sin\theta \simeq0.39$ as an example, the $\Xi_{bc}^+$ branching ratio would be larger by a factor of approximately 4 than $\Xi_{bc}^0$, and the detection efficiency at the LHCb would also be improved by a similar factor owing to more displaced decay vertices~\cite{LHCb:2020iko}, so it is expected that approximately 16 times events can be collected by using $\Xi_{bc}^+$ decays than using a similar $\Xi_{bc}^0$ decay. Therefore, we propose that $\Xi_{bc}^+\to D^{(*)+}\Lambda$, $\Xi_{bc}^+\to \Lambda_c^+\bar{K}^{(*)0}$,  $\Xi_{bc}^+\to D^{(*)+}K^-p$, and $\Xi_{bc}^+\to D^{0}\bar{K}^{(*)0}p$ should be searched experimentally with priority. They can be searched in the $p2\pi^+2\pi^-$, $p2\pi^+\pi^-K^-$ and $p2\pi^+2K^-$ final states.

\section{Conclusion}

We have calculated the inclusive $\Xi_{bc}^{+,0}\to X_{cs}$ decay widths making use of the local OPE technique in the framework of HDET. The $\Xi_{bc}$ is treated as the superposition of $\Xi_\mathcal{S}$ containing a scalar $bc$ diquark and $\Xi_\mathcal{X}$ containing an axial-vector $bc$ diquark. The mixing angle $\theta$ dependent results for both the $\Xi_{bc}^{+,0}$ lifetimes and the inclusive $\Xi_{bc}^{+,0}\to X_{cs}$ branching ratios have been obtained. The corresponding $\Xi_{bc}^{+,0}\to X_{cd}$ results are also presented. It turns out that these observables are very sensitive to $\theta$, and thus their measurements can help determine the $bc$ diquark spin property in $\Xi_{bc}^{+,0}$. With the reference mixing angle $\sin\theta$ = 0.39~\cite{Roberts:2008wq}, the $\Xi_{bc}^+$ lifetime is approximately four times longer than that of $\Xi_{bc}^0$, and the branching ratio $\mathcal{B}(\Xi_{b c}^{+} \to  X_{cs})$ is approximately four times bigger than $\mathcal{B}(\Xi_{b c}^0 \to  X_{cs})$. We propose that the decay channels $\Xi_{bc}^+\to D^{(*)+}\Lambda$, $\Xi_{bc}^+\to \Lambda_c^+\bar{K}^{(*)0}$,  $\Xi_{bc}^+\to D^{(*)+}K^-p$, and $\Xi_{bc}^+\to D^{0}\bar{K}^{(*)0}p$ should be used in $\Xi_{bc}$ searches. 

{\it Acknowledgement.} --- This work is supported by the Natural Science Foundation of China under grant No. 12005068.

\begin{appendix}

\section{Appendix: Weak-annihilation contributions to $\Xi_{bc}$ total widths}
\label{app}

The transition operators in the $bq$ and $cq$ weak-annihilation contributions to the $\Xi_{bc}$ decay widths are
\begin{eqnarray}
	\mathcal{T}^{an}_{bq}
	=\frac{G_{F}^{2}\left|V_{c b}\right|^{2}}{2 \pi m_{bu}^{2}} \left(m_{bu}^{2}-m_{c}^{2}\right)^{2}\left(C_{1}-C_{2}\right)^{2}\nonumber\\
\left(\bar{b}\gamma_{\mu}(1-\gamma_{5}) b \right)\left(\bar{q}\gamma^{\mu}(1-\gamma_{5}) q\right),\\
	\mathcal{T}^{an}_{cq}
	=\frac{G_{F}^{2}\left|V_{c s}\right|^{2}}{2 \pi m_{cd}^{2}} \left(m_{cd}^{2}-m_{s}^{2}\right)^{2}\left(C_{1}-C_{2}\right)^{2}\nonumber\\
\left(\bar{c}\gamma_{\mu}(1-\gamma_{5}) c \right)\left(\bar{q}\gamma^{\mu}(1-\gamma_{5}) q\right), 
\end{eqnarray}
where the diquark masses $m_{bu}$ and $m_{cd}$ are approximately the heavy quark mass $m_b$ and $m_c$ respectively. 
The hadronic matrix elements of these operators are estimated in the non-relativistic potential model~\cite{Cheng:2019sxr} as, 
\begin{eqnarray}	
&	&\left\langle\Xi_{\mathcal{S}}^{+}\left|\left(\bar{b}\gamma_{\mu}(1-\gamma_{5}) b \right)\left(\bar{q}\gamma^{\mu}(1-\gamma_{5}) q\right)\right| \Xi_{\mathcal{S}}^{+}\right\rangle \nonumber\\
&	=&\left\langle\Xi_{\mathcal{S}}^{0}\left|\left(\bar{c}\gamma_{\mu}(1-\gamma_{5}) c \right)\left(\bar{q}\gamma^{\mu}(1-\gamma_{5}) q\right)\right| \Xi_{\mathcal{S}}^{0}\right\rangle \nonumber\\
&	=&2 m_{bc} \left|\psi_{q,bc}(0)\right|^{2},\\
&	&\left\langle\Xi_{\mathcal{X}}^{+}\left|\left(\bar{b}\gamma_{\mu}(1-\gamma_{5}) b \right)\left(\bar{q}\gamma^{\mu}(1-\gamma_{5}) q\right)\right| \Xi_{\mathcal{X}}^{+}\right\rangle \nonumber\\
&	=&\left\langle\Xi_{\mathcal{X}}^{0}\left|\left(\bar{c}\gamma_{\mu}(1-\gamma_{5}) c \right)\left(\bar{q}\gamma^{\mu}(1-\gamma_{5}) q\right)\right| \Xi_{\mathcal{X}}^{0}\right\rangle \nonumber\\
&	=&6 m_{bc} \left|\psi_{q,bc}(0)\right|^{2},\\
&	&\left\langle\Xi_{\mathcal{S}}^{+}\left|\left(\bar{b}\gamma_{\mu}(1-\gamma_{5}) b \right)\left(\bar{q}\gamma^{\mu}(1-\gamma_{5}) q\right)\right| \Xi_{\mathcal{X}}^{+}\right\rangle \nonumber\\
	&=&	\left\langle\Xi_{\mathcal{S}}^{0}\left|\left(\bar{c} \gamma_{\mu}(1-\gamma_{5})c \right)\left(\bar{q}\gamma^{\mu}(1-\gamma_{5}) q\right)\right| \Xi_{\mathcal{X}}^{0}\right\rangle \nonumber\\
	&=&2\sqrt{3} m_{bc} \left|\psi_{q,bc}(0)\right|^{2}.
\end{eqnarray}
Then, the leading-order weak-annihilation inclusive decay widths are given by 
\begin{eqnarray}		
&\Gamma_{\mathcal{S}}^{bq}
	=\frac{G_{F}^{2}\left|V_{c b}\right|^{2}}{2 \pi m_{bu}^{2}}|\psi_{q,bc}(0)|^2 \left(m_{bu}^{2}-m_{c}^{2}\right)^{2}(C_{1}-C_{2})^{2},\nonumber\\
&\Gamma_{\mathcal{S}}^{cq}
	=\frac{G_{F}^{2}\left|V_{c s}\right|^{2}}{2 \pi m_{cd}^{2}}|\psi_{q,bc}(0)|^2 \left(m_{cd}^{2}-m_{s}^{2}\right)^{2}(C_{1}-C_{2})^{2},\nonumber\\
&\Gamma_{\mathcal{X}}^{bq}=3\Gamma_{\mathcal{S}}^{bq},\Gamma_{\mathcal{SX}}^{bq}=\sqrt{3}\Gamma_{\mathcal{S}}^{bq}\nonumber\\
&\Gamma_{\mathcal{X}}^{cq}=3\Gamma_{\mathcal{S}}^{cq},\Gamma_{\mathcal{SX}}^{cq}=\sqrt{3}\Gamma_{\mathcal{S}}^{cq}.
\end{eqnarray}

\end{appendix}

\end{document}